\documentclass[conference,compsoc]{IEEEtran}
\IEEEoverridecommandlockouts
\ifCLASSOPTIONcompsoc
  \usepackage[nocompress]{cite}
\else
  \usepackage{cite}
\fi
\ifCLASSINFOpdf
\else
\fi
\usepackage{booktabs}
\usepackage{multirow} 
\usepackage{diagbox}
 
\usepackage{amsmath,amsthm,amssymb,amsfonts}
\usepackage{algorithm}
\usepackage{algorithmic}
\usepackage{listings}
\usepackage{amsthm}
\usepackage{graphicx}
\usepackage{subcaption}
\usepackage{tabularray}
\usepackage[dvipsnames]{xcolor}
\usepackage{caption}
\usepackage{float} 
\usepackage{tcolorbox}

\lstdefinestyle{Java}{
	language        =   Java,
	basicstyle      =   \zihao{-5}\ttfamily,
	numberstyle     =   \zihao{-5}\ttfamily,
	keywordstyle    =   \color{blue},
	keywordstyle    =   [2] \color{teal},
	stringstyle     =   \color{magenta},
	commentstyle    =   \color{red}\ttfamily,
	breaklines      =   true,  
	columns         =   fixed,  
	basewidth       =   0.5em,
}

\usepackage{marvosym}

\begin{document}
%
\title{LAMD: Context-driven Android Malware Detection and Classification with LLMs}




%
\author{\IEEEauthorblockN{Xingzhi Qian\IEEEauthorrefmark{1}\IEEEauthorrefmark{2},
Xinran Zheng\IEEEauthorrefmark{1}\IEEEauthorrefmark{2},
Yiling He\textsuperscript{\Letter}\IEEEauthorrefmark{2}, 
Shuo Yang\IEEEauthorrefmark{3} and
Lorenzo Cavallaro\IEEEauthorrefmark{2}}
\IEEEauthorblockA{\IEEEauthorrefmark{2}University College London\quad\IEEEauthorrefmark{3}University of Hong Kong}
\IEEEauthorblockA{\IEEEauthorrefmark{2}\{xingzhi.qian.23, xinran.zheng.23, yiling-he, l.cavallaro\}@ucl.ac.uk}
\IEEEauthorblockA{\IEEEauthorrefmark{3}shuoyang.ee@gmail.com}
\thanks{*Equal contribution and co-first authors. \Letter Corresponding Author.}
}


\maketitle

\begin{abstract}
The rapid growth of mobile applications has escalated Android malware threats. Although there are numerous detection methods, they often struggle with evolving attacks, dataset biases, and limited explainability. Large Language Models (LLMs) offer a promising alternative with their zero-shot inference and reasoning capabilities. However, applying LLMs to Android malware detection presents two key challenges: (1)~the extensive support code in Android applications, often spanning thousands of classes, exceeds LLMs' context limits and obscures malicious behavior within benign functionality; (2)~the structural complexity and interdependencies of Android applications surpass LLMs' sequence-based reasoning, fragmenting code analysis and hindering malicious intent inference. To address these challenges, we propose LAMD, a practical context-driven framework to enable LLM-based Android malware detection. LAMD integrates key context extraction to isolate security-critical code regions and construct program structures, then applies tier-wise code reasoning to analyze application behavior progressively, from low-level instructions to high-level semantics, providing final prediction and explanation. A well-designed factual consistency verification mechanism is equipped to mitigate LLM hallucinations from the first tier. Evaluation in real-world settings demonstrates LAMD's effectiveness over conventional detectors, establishing a feasible basis for LLM-driven malware analysis in dynamic threat landscapes.

\end{abstract}


%
\IEEEpeerreviewmaketitle

\section{Introduction}
The rapid expansion of the Android ecosystem has heightened security risks, with malware posing serious threats to user privacy, financial security, and sensitive data. Over the past decade, researchers have developed various Android malware detection techniques, yet these methods face persistent challenges in real-world scenarios. 
Firstly, the open and evolving nature of Android complicates the detection of adaptive malware~\cite{transcend, transcending}. 
Reliance on specific datasets introduces biases, such as ambiguous timestamps and randomly selected samples~\cite{tesseract}, which further compromise model reliability. 
Additionally, conventional detectors often lack explainability that fail to offer clear, human-readable insights into malicious behaviors.

Large Language Models (LLMs) offer a promising paradigm shift in malware detection, differing fundamentally from conventional detectors. They achieve zero-shot inference relying on vast pre-trained knowledge instead of specifically labelled datasets~\cite{large_zero,toolformer}, allowing them to handle the evolving malware and potential training bias. 
Furthermore, to bridge the gap of explainability, the advanced generative capabilities of LLMs present an opportunity by providing human-readable comprehension, thereby enhancing malware analysis from both an accuracy and interpretability perspective.


\begin{figure}[t]
    \centering
    \includegraphics[width=0.85\linewidth]{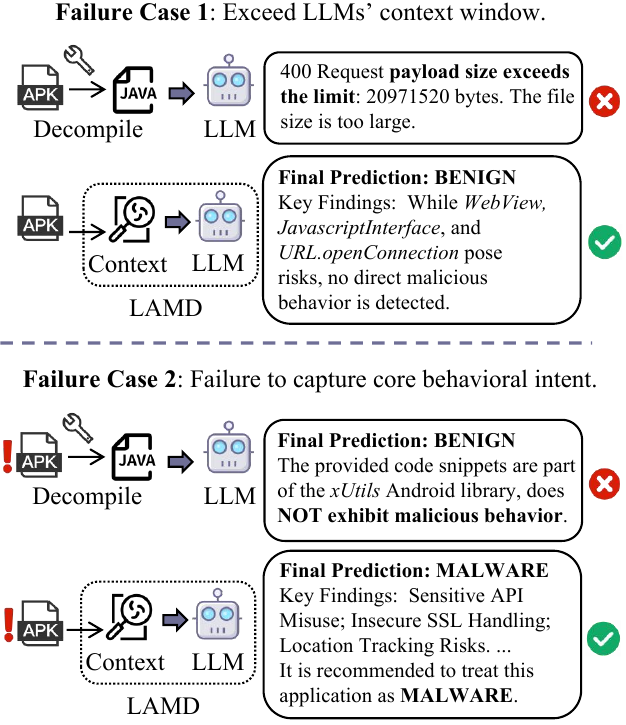}
    \caption{Two failure cases of applying LLMs to Android malware detection: (1) context window limitations and (2) failure to capture malicious intent. LAMD addresses these challenges by extracting key contexts and tiered reasoning to capture structures and semantics efficiently.}
    \label{fig:motivation}
\end{figure}


However, despite the potential, LLMs are not omnipotent. Two primary challenges hinder their effectiveness in Android malware detection: \textit{\textbf{(1) Excessive support codes in Android applications}}: Android malware often comprises thousands of classes to support diverse functionalities across various devices. 
While some LLMs support up to 2 million tokens~\cite{gemini1.5}, directly processing decompiled malware remains impractical, as some samples would still exceed this limit.
Truncation offers a potential alternative, but it can cause substantial contextual loss, ultimately degrading analysis accuracy~\cite{truncate}.
More importantly, these normal functional codes make malicious codes sparse within the program and become extremely hard to detect.
\textit{\textbf{(2) Complex program structures}}: Although code exhibits structural characteristics akin to natural language~\cite{naturalness}, it remains fundamentally distinct due to inherent structural complexity~\cite{codeandnatural}. 
For instance, deeply nested dependencies, intricate API interactions, and class hierarchies all extend beyond sequential token-based modeling. 
The complexity is further amplified in Android malware, where obfuscation techniques, multi-component interactions, and convoluted function invocations obscure malicious intent. 
At its core, these challenges raise a fundamental research question: 
\begin{center}
\fcolorbox{white}{gray!10}{\parbox{.9\linewidth}{\textit{Can we extract crucial semantic and structural information from complete application to guide LLMs in detecting Android malware?}}}
\end{center}

To address this, we draw inspiration from how analysts examine Android malware in the real world, where they identify suspicious APIs, interfaces, and function calls, analyzing contextual relationships to detect malicious behavior within extensive application code. 
Our framework aims to guide LLMs in replicating this analytical process, enhancing automated explainable Android malware detection. 

We propose LAMD, a novel and practical framework that enables LLMs for explainable Android malware detection. LAMD consists of two core components: key context extraction and tier-wise code reasoning. Specifically, we perform static analysis on APKs and employ a custom backward slicing algorithm to extract key variables, dependencies, and invocations for predefined suspicious APIs. These elements are transformed into graphical representations, filtering irrelevant code and preserving essential semantics for the LLM. 
Furthermore, we introduce tier-wise code reasoning combined with factual consistency verification, transferring structured knowledge across multiple tiers to understand, draw conclusions and explain detection results. 
This approach refines the LLM’s understanding of applications, progressing from fine-grained analysis to higher-level abstraction. 
As shown in Figure~\ref{fig:motivation}, LAMD effectively addresses real-world challenges, enabling practical and explainable LLM-powered Android malware detection.
The main contributions of this paper are as follows:

\begin{itemize}
    \item We introduce LAMD, the first LLM-powered practical Android malware detection framework, unlocking LLMs' ability for explainable Android malware detection in dynamic scenarios, providing heuristics for LLM-powered malware-related tasks.
    \item LAMD integrates key context extraction and tier-wise code reasoning to filter irrelevant functionalities while capturing semantics and structural dependencies. A targeted factual consistency verification strategy is also established to ensure accurate reasoning. 
    \item We evaluate LAMD on a collected dataset\footnote{
    The dataset is open-source for further research on LLM-based malware tasks: https://doi.org/10.5281/zenodo.14884736
    } 
    that reflects the real-world setting. Our results show that LAMD outperforms conventional learning-based detectors under distribution shift, demonstrating its effectiveness in detecting and explaining fast evolving Android malware. Furthermore, we discuss the tension between learning-based models and pre-trained LLMs, highlighting promising directions for future research.
\end{itemize}

\section{Related Work}
This section reviews conventional Android malware detection, their real-world limitations, and recent advances in LLM-powered Android malware detection, situating our work in this evolving field.

\subsection{Learning-based Android Malware Detection}
Learning-based Android malware detectors leverage machine learning or deep learning models to automatically learn patterns from features extracted by static or dynamic analysis, enhancing scalability and adaptability~\cite{unraveling}. Due to the high cost of dynamic analysis, most models rely on static feature extraction through reverse engineering~\cite{Arpdrebin, Grossedeepdrebin}. Despite advancements, these models struggle with concept drift, where evolving malware variants degrade performance in real-world deployments~\cite{transcend, transcending}. Current works try to mitigate it, some of them focus on exploring robust features~\cite{scrr,damo,svm_ce,apigraph} and others leverage continual learning~\cite{continuous, tesseract,malware_evolution_update} or active learning~\cite{droidevolver, online_mal} to let models adapt to new distribution. However, these methods either target to specific feature space~\cite{scrr} or introduce high retraining overhead and the risk of label poisoning~\cite{recda, labelless}. An additional challenge is explainability, which is critical for security analysis. Existing methods primarily use feature attribution techniques, providing importance scores without generating human-readable behavioral analysis~\cite{finer, belaid}. To address these limitations, researchers are increasingly exploring LLM-based approaches, which leverage extensive external knowledge and reasoning capabilities~\cite{code_explain,code_explain_2}, while tasks related to malware detection and analysis still remain under-explored.

\begin{figure*}[t]
    \centering
    \setlength{\abovecaptionskip}{0cm}
    \includegraphics[width=1.0\linewidth]{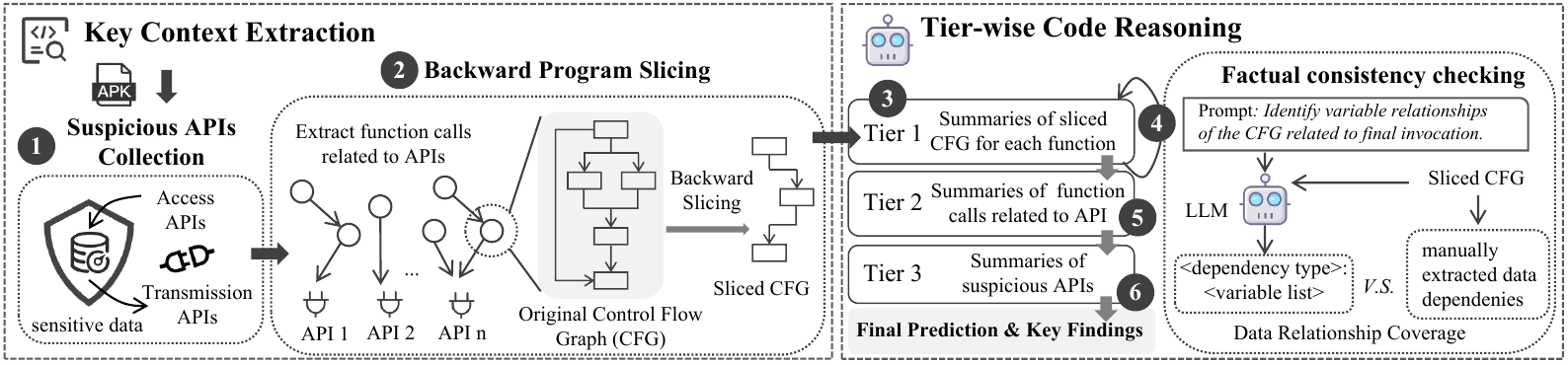}
    \caption{The workflow of LAMD. Suspicious APIs are identified via predefined rules (Step 1), and their calling functions with control flow graphs are extracted through static analysis. A customized backward slicing technique refines relevant instructions, preserving potential malicious intent (Step 2). In the code reasoning phase, the structured control flow graph, function relationships, and suspicious APIs form hierarchical tiers for malware detection and human-readable explanations (Steps 3-6). Factual consistency verification ensures first-tier summary reliability, mitigating hallucination (Step 4).}
    \label{fig:arch}
\end{figure*}

\subsection{LLM-powered Malware Detection}

LLMs are increasingly used in security tasks like code analysis~\cite{code_sum}, vulnerability detection~\cite{llm4vuln,llm4vuln_bench}, and malware classification~\cite{llm_syscall}. Unlike traditional learning-based models, LLMs offer zero-shot inference capabilities, enabling them to generalize beyond predefined training data and feature spaces~\cite{large_zero,toolformer}. This adaptability leads researchers to explore LLMs for malware detection and analysis, demonstrating their potential in security tasks. However, most studies focus on relatively simple malware ecosystems, such as npm packages~\cite{npm_malware}, PowerShell scripts~\cite{raconteur}, Linux binaries~\cite{llm_syscall}, and JavaScript-based threats~\cite{tactics}.

Initial attempts at LLM-based Android malware detection are limited. Walton \textit{et al.}~\cite{malware_exploring} proposed a hierarchical approach, analyzing decompiled code at the function, class, and package levels. However, the lack of filtering mechanisms allows benign code to obscure malicious patterns, reducing detection accuracy and increasing computational costs. Even with a balanced 200-sample dataset, their best prompt only achieved 75\% accuracy. Zhao \textit{et al.}~\cite{apppoet} relies on predefined feature spaces (Drebin~\cite{Arpdrebin}) to generate feature summaries instead of analyzing raw code. These summaries are then embedded and fed into a deep neural network (DNN) for training and detection. While focusing on the feature's name is efficient, it lacks structural and invocation insights. Additionally, as a learning-based method, it also inherits generalization issues and dataset bias influences of conventional models. These limitations highlight the need for a framework that leverages LLMs effectively in real-world Android malware detection, which integrates both structural and semantic context.

\section{Methodology}
This section outlines the core components of our framework, LAMD, and how they cooperate to detect and understand Android malware efficiently.

\subsection{Overall Architecture}
The LAMD framework is designed to extract essential functionalities and their contextual information, enabling LLMs to generate both detection and reasoning results. It consists of two key components: (1) Key Context Extraction and (2) Tier-wise Code Reasoning, detailed as follows:
\begin{itemize}
    \item \textbf{Key Context Extraction}: This module identifies suspicious APIs as seed points and analyzes their control and data dependencies within the application. It provides a structured representation of key program behaviors by pruning the calling relationships of potentially malicious interactions.
    \item \textbf{Tier-wise Code Reasoning}: To preserve contextual integrity while managing token limitations, LAMD employs a tiered reasoning strategy. It processes information at three levels—function, API, and APK—where the output of each tier informs the next. To mitigate error propagation, factual consistency verification is applied at the first tier.
\end{itemize}
Figure~\ref{fig:arch} shows the pipeline of our framework. Overall, the raw input to LAMD consists of APK files, from which suspicious APIs and their sliced contexts are extracted as the input for LLMs. The malicious behavior of the application is then determined through three tiers of code reasoning.

\subsection{Key Context Extraction}
\subsubsection{Suspicious API Collection}
Malware exploits system vulnerabilities or API permissions to steal data, manipulate resources, or maintain persistence. Many attacks rely on sensitive API calls to implement malicious behaviors. We perform static analysis on APKs to extract suspicious APIs as key context to identify malware. Let $\mathcal{A} = \{a_1, a_2, \ldots, a_n\}$ be the set of all API calls in an APK. A subset $\mathcal{A}_{sus} \subset \mathcal{A}$ is deemed suspicious if it interacts with sensitive components, executes malware-associated operations or exposes sensitive data. These APIs fall into two categories:
\begin{itemize}
    \item \textit{Sensitive data access APIs}: Many apps handle sensitive data, but assessing developer trustworthiness is challenging. Monitoring APIs accessing such data is crucial. Smartphone OSs enforce permission-based access control, requiring declared permissions for some APIs, while others, like \verb|getPrimaryClip()|, bypass enforcement. Therefore, an API $a_i$ falls into this category if it either: (1) requires explicit permissions for access control, or (2) grants direct access to sensitive user data without permission enforcement.
    \item \textit{Sensitive data transmission APIs}: Monitoring potential data exfiltration channels is critical, as malware often exploits these APIs—commonly referred to as sink APIs—to transmit sensitive information to external entities. An API $a_j$ is classified as suspicious if it facilitates the transfer of sensitive data to an external environment.
\end{itemize}

To extract suspicious APIs, we leverage publicly available knowledge based on PScout~\cite{pscout}, SuSi~\cite{susi} and Flowdroid~\cite{flowdroid} to label them. In a real-world setting, not every application contains suspicious APIs. These samples should be treated individually, where traditional learning-based malware detectors also struggle to handle~\cite{he2022msdroid}.

\subsubsection{Backward Program Slicing}

Extracting suspicious APIs is useful, but analyzing them in isolation often obscures their context. For instance, while \verb|sendTextMessage()| is legitimate in messaging apps, malware may exploit it for premium-rate SMS. To capture intent, we extract functions invoking suspicious APIs and refine their control flow graphs (CFGs), $G = <N, E>$, where nodes $n \in N$ represent basic blocks or instructions, and edges $(e_1, e_2) \in E$ define control flow. 

Recent work reveals that decoder-only Transformers, at the core of modern LLMs like GPT-4o-mini, are susceptible to over-squashing~\cite{barbero2024transformers}, where information from distant tokens is compressed and loses influence. The authors show how this may impair models in learning proper representations of similar long sequences, leading to representation clashes and wrong outcomes for downstream tasks. The study suggests that input-compressing strategy might preserve a high the signal-to-noise ratio, able to improve performance on downstream tasks. We follow this intuition in our context (the analysis of large complex programs), aiming at providing quality context to the models to improve on the downstream tasks. Since CFGs can be large and noisy, containing instructions that are irrelevant to suspicious API usages, directly inputting them to LLM risks conflating malicious and benign patterns, thereby obscuring sensitive behaviors. We therefore apply backward slicing~\cite{slicing} to isolate instructions affecting the API invocation and provide meaningful context to LLM. A slice $S$ is defined by a slicing criterion $C = <s, V>$ where $s$ is the statement invoking $a_i$ and $V$ includes all parameters. We classify relevant variables as: (1) \textit{Direct relevant variables}: Variables' values can affect variable $v \in V$ of $a_i$  (2) \textit{Indirect relevant variables}: Variables in branch statements whose value affects invocation of $a_i$. The backward slicing is to select the set of instructions in $\mathcal{P}$ that directly or indirectly affect the execution or parameters of $a_i$. The backward slicing algorithm consists of two steps to ensure completeness in complex branch structures:
\begin{itemize}
    \item \textbf{Variable retrieval}: Identify all variables contributing to the parameters (and internal states) used by the suspicious API and store them in a candidate set.
    \item \textbf{Slices extraction}: Append instructions related to variables collected in the first step.
\end{itemize}

After slicing, we generate sliced CFGs for each sensitive API, preserving essential control flow and statements. Notably, If undeclared variables remain in a sliced function, inter-procedural backward slicing is recursively applied to its callers until all variables are resolved. The details of slicing algorithm are shown in Appendix~\ref{slicing}.

\subsection{Tier-wise Code Reasoning}
Code reasoning involves analyzing and interpreting code to understand its behavior, identify potential threats, and generate meaningful explanations. We propose a three-tier reasoning strategy that refines APK behavior analysis from fine- to coarse-grained levels, enhancing both prediction accuracy and interpretability. This hierarchical approach improves malicious component identification, mitigates LLM token-length limitations, and captures structural and invocation semantics through separable reasoning, ensuring a more effective and scalable malware detection framework.


\subsubsection{Tier 1: Function Behavior Summarization}
In the previous stage, several functions invoking suspicious APIs are extracted and sliced to maintain the related context. Each sliced CFG of the function is fed to LLM to capture low-level code patterns and functionalities. 

\begin{tcolorbox}[title=Tier 1: Function Behavior Summarization Prompt, 
left=2pt, 
right=2pt, 
top=3pt, 
bottom=3pt, 
fonttitle=\small,colback=gray!20, colframe=black, colbacktitle=black, coltitle=white, sharp corners, fontupper=\small, fontlower=\small, before upper=\raggedright, before lower=\raggedright]
You are a cybersecurity expert specializing in Android malware analysis. Analyze the provided control flow graph including instructions related to sensitive API calls in detail.

Control Flow Graph: $\{CFG\_content\}$
\end{tcolorbox}

\subsubsection{Tier 2: API Intent Awareness}
The context of a specific API is typically determined by a series of functions. Beyond function invocation relationships, it is crucial to examine how inter-function associations influence the API's intent. Due to diverse contexts, an API may appear in multiple Function Call Graphs (FCGs). For instance, \verb|getDeviceId()| is benign when used solely for local logging but becomes malicious when invoked within \verb|sendImeiToServer()|, where it exfiltrates the IMEI to a remote server. Therefore, at this mid-tier, all functions associated with a suspicious API are structured into multiple FCGs to analyze its overall intent. Each node in an FCG is represented by the generated function summary in tier 1. 

\begin{tcolorbox}[title=Tier 2: API Intent Awareness Prompt, 
left=2pt, 
right=2pt, 
top=3pt, 
bottom=3pt, 
fonttitle=\small,colback=gray!20, colframe=black, colbacktitle=black, coltitle=white, sharp corners, fontupper=\small, fontlower=\small, before upper=\raggedright, before lower=\raggedright]
You are a cybersecurity expert specializing in Android malware analysis. Analyze the main functionality and behavior of the provided sensitive API based on the function call graphs and a summary of each function's behavior.

API name: $\{API\ name\}$ \\
API type: $\{access/transfer\}$\\
for $i$-th Function Call Graph: \\
\makebox[2em]{} FCG: $\{FCG\_content\}$ \\
\makebox[2em]{} $\{function\_name\} \{function\_summary\}$ 
\end{tcolorbox}

\subsubsection{Tier 3: APK Maliciousness Judgement}
After extracting intents from suspicious APIs in an APK, LLMs assess its maliciousness and justify their decision. and generate Indicators of Compromise (IoCs), summarizing sensitive data access, external transmissions, and anomalous behavior to enhance transparency and trust.


\begin{tcolorbox}[title=Tier 3: APK Maliciousness Judgement Prompt, 
left=2pt, 
right=2pt, 
top=3pt, 
bottom=3pt, 
fonttitle=\small,colback=gray!20, colframe=black, colbacktitle=black, coltitle=white, sharp corners, fontupper=\small, fontlower=\small, before upper=\raggedright, before lower=\raggedright]
You are a cybersecurity expert specializing in Android malware analysis. Determine whether the application is MALWARE or BENIGN, citing indicators of compromise, evidence, and malicious patterns if present. Give a final prediction and key findings of your analysis.

for $i$-th API: \\
\makebox[2em]{}API name: $\{API\ name\}$ \\
\makebox[2em]{}API type: $\{access/transfer\}$\\
\makebox[2em]{}API intent: $\{API\ summary\}$
\end{tcolorbox}


\subsubsection{Factual Consistency Verification}
Generating behavior summaries with LLMs risks hallucinations~\cite{nature24hallucinations}, producing facts inconsistent with instructions. To prevent error accumulation, we verify function-level summaries before higher-tier reasoning, leveraging their limited dependencies and concise structure.

Building on factual consistency verification~\cite{cloze, factasking}, we design a structured template to capture data dependencies in sliced CFGs. To enhance inference, we prompt the LLM to select corresponding data relationships from the input function based on specific definitions. We define five dependencies: variable-to-API interactions (direct, transitive, conditional) and inter-variable relationships (parallel, derived). The former tracks how variables influence API execution via assignments, call chains, and control flow, while the latter captures joint computation and derivation. Loop dependencies are excluded, as Soot expands loops when analysing binary code. Appendix~\ref{datadependency} details these dependencies. For consistency verification, we integrate an in-context learning-based prompt with function summarization, querying the LLM to extract dependencies in the format: $<dependencies\ type>:<variable\ names>$.

\begin{tcolorbox}[title=Factual Consistency Verification Prompt, 
left=2pt, 
right=2pt, 
top=3pt, 
bottom=3pt, 
fonttitle=\small,
colback=gray!20, colframe=black, colbacktitle=black, coltitle=white, sharp corners, fontupper=\small, fontlower=\small, before upper=\raggedright, before lower=\raggedright]
The provided control flow graph represents a slice of the function, identifying variable relationships for each statement leading to the final invocation statement that invokes $\{function\_name\}$. The output should follow the template: $\{template\}$

Control Flow Graph: $\{CFG\_content\}$ 

There are FIVE types of relationships: \\
1. Direct: Variables used directly as function parameters.\\
\textit{Example}: $invoke r1.method(r2) \rightarrow {r1, r2}$ \\
2. Transitive: Variables whose values flow through assignments but are not directly used in the invocation. \\
\textit{Example}: $r2 = r3.getValue()$; \\
$invoke r1.method(r2) \rightarrow {r3}$ \\
...
\end{tcolorbox}

To evaluate the reliability of generated summaries, we propose a Data Relationship Coverage (DRC) metric:
\begin{equation}
DRC = \frac{\#\{\text{correctly completed dependencies}\}}{\#\{\text{all selected dependencies}\}}.
\end{equation}

A summary is considered factually consistent if the LLM accurately reconstructs variable dependencies, i.e., $DRC \ge \theta$, where $\theta$ is a reliability threshold. Otherwise, the summary is revised to mitigate inaccuracies.


\section{Evaluation}

\begin{table}
\centering
\caption{Overall detection performance (\%) across three test sets with increasing distribution drift (Test 1 $<$ Test 2 $<$ Test 3).}
\label{tab:detection}
\setlength{\tabcolsep}{1.8pt}
\begin{tabular}{lccccccccc}\toprule
\multirow{2}{*}{\textbf{Model}} & \multicolumn{3}{c}{\textbf{Test 1}} & \multicolumn{3}{c}{\textbf{Test 2}} & \multicolumn{3}{c}{\textbf{Test 3}} \\
\cmidrule(lr){2-4} \cmidrule(lr){5-7} \cmidrule(lr){8-10}
                         & F1   & FPR          & FNR   & F1   & FPR          & FNR   & F1   & FPR          & FNR   \\\midrule
Drebin~\cite{Arpdrebin}  & 81.33 & \textbf{0.40} & 24.21 & 73.60 & 0.99          & 31.14 & 61.59 & 4.36          & 37.00 \\
DeepDrebin~\cite{Grossedeepdrebin} & 71.92 & 0.62 & 34.12 & 69.22 & 0.85          & 36.13 & 66.11 & 0.95          & 38.58 \\
Malscan~\cite{malscan}   & 66.37 & 0.73          & 46.83 & 65.45 & \textbf{0.84} & 47.73 & 64.54 & \textbf{0.91} & 48.30 \\
LAMD-R                   & 75.34 & 5.39          & 15.24 & 75.71 & 4.79          & 12.99 & 75.83 & 4.84          & 11.19 \\
LAMD-F                   & 87.63 & 2.00          & 10.37 & 87.28 & 1.85          & 9.74  & 87.21 & 1.77          & 9.83  \\
LAMD                     & \textbf{90.24} & 1.26 & \textbf{8.44} & \textbf{90.16} & 1.38 & \textbf{7.79} & \textbf{89.85} & 1.30 & \textbf{8.47} \\
\bottomrule
\end{tabular}
\end{table}
This section presents a comprehensive evaluation of LAMD, assessing its Android malware detection performance in real-world scenarios and the quality of its generated explanations through a series of experiments. 

Our experiments are conducted on an RTX A6000 GPU. For LLM-based reasoning, we utilize GPT-4o-mini~\cite{openai}, selected for its high efficiency, cost-effectiveness, and strong reasoning capabilities. The Data Relation Coverage (DRC) threshold $\theta$ is set to 0.95 to balance computational efficiency and accuracy.



\subsection{Dataset Construction}
To ensure a realistic dataset, we adhere to the following principles~\cite{tesseract}: (1) Maintain temporal order in training and testing; (2) Preserve the real-world malware-to-benign ratio; (3) Ensure diversity by including packed, obfuscated, and varied market samples.

Based on these principles, we select Android APKs from Androzoo~\cite{AndroZoo} according to their discovery time, which is determined by their submission to VirusTotal~\cite{VirusTotal} (as release timestamps can be unreliable\footnote{Due to modifiable or randomly generated release times, some timestamps (dex date) are inaccurate.}). 
The dataset includes a decade of benign and malicious samples, labeled based on VirusTotal analysis from Androzoo. Samples flagged by more than four vendors are classified as malicious. To track malware family evolution, we used Euphony~\cite{euphony} to extract family labels.
The dataset spans from 2014 to 2023\footnote{as APK labels generally stabilize after about one year in the wild~\cite{labelstability}, we choose samples before 2024}, comprising 13,794 samples. 
The remaining samples are evenly split into three test sets (about 3,015 each) to represent incremental distribution drift.
Table~\ref{tab:dataselection} shows details about the used dataset.

\subsection{Metrics}
\textbf{(1) Classification Metrics.} To address class imbalance in Android malware datasets, we use the F1-score to balance precision and recall, while also minimizing False Positive Rate (FPR) and False Negative Rate (FNR) to improve accuracy and reduce manual analysis overhead. Results are reported as percentages. \textbf{(2) Summarization Metrics.} Effective malware analysis provides interpretable insights and aids manual audits. Since malware family identification often requires expert review, initial categorization prioritizes the common sense of behavior patterns. Following prior work~\cite{malcategory}, we mainly consider six categories: Adware, Backdoor, PUA (Potentially Unwanted Applications), Riskware, Scareware, and Trojan. For evaluation, we adopt a ChatGPT-based metric~\cite{llm4codeanalysis, gptmetric1, gptmetric2}, where GPT-4o-mini~\cite{openai} assesses whether LAMD’s detection aligns with the expected behaviors of each category~\cite{humaneval}.

\subsection{Baselines}
In Android malware detection, an APK serves as input, containing the codebase (e.g., .dex files) and configuration files (e.g., AndroidManifest.xml), which provide behavioral insights like API calls and permissions, represented in vector or graph formats. We evaluate LAMD against Drebin~\cite{Arpdrebin}, DeepDrebin~\cite{Grossedeepdrebin}, and Malscan~\cite{malscan} which are representative learning-based methods on these feature formats, with details in Appendix~\ref{baseline}. To assess component impact, LAMD-R removes tier-wise reasoning to test structural and semantic analysis, while LAMD-F retains reasoning but omits factual consistency verification to evaluate hallucination control.


\begin{table}
\centering
\caption{Out-of-distribution (OOD) detection performance (\%) on post-cutoff samples for LAMD and aligned training-temporal scope for learning-based models.}
\label{tab:ood_detection}

\begin{tabular}{lccc}\toprule
\textbf{Model} & {F1} & {FPR}          & {FNR} \\\midrule
Drebin~\cite{Arpdrebin}            & 83.72 & 0.42          & 22.02 \\
DeepDrebin~\cite{Grossedeepdrebin} & 79.78 & 0.39          & 26.80 \\ 
Malscan~\cite{malscan}             & 83.87 & {\textbf{0.00}} & 46.77 \\
LAMD                               & {\textbf{85.71}} & 1.89 & {\textbf{14.29}} \\
\bottomrule
\end{tabular}
\end{table}

\subsection{Experiments}
We conduct a series of experiments to thoroughly evaluate LAMD's performance in malware detection and the quality of its generated explanation.

\subsubsection{Malware Detection Performance}
Table~\ref{tab:detection} compares LAMD’s detection performance with baselines on test datasets with the increasing level of drift. LAMD improves F1-scores by 23.12\% and reduces FNR by 71.59\% on average, enhancing detection reliability. While the FPR shows a slight increase, it is less indicative of true performance due to class imbalance, where learning-based methods often misclassify malware as benign because of the dominance of benign samples. The performance drop in LAMD-R highlights the necessity of hierarchical code summarization, and the slight decline in LAMD-F underscores the role of hallucination mitigation. By refining input code and extracting key information, LAMD minimizes hallucination risks, ensuring more reliable malware detection. 



We performed an additional out-of-distribution (OOD) experiments designed to simulate a temporal natural drift by aligning GPT-4o-mini's pre-training cutoff date (October 2023) with a dataset comprising samples of newer timestamps. As detailed in Table~\ref{ood_datasetdescription}, to make a fair comparison, LAMD is tested on samples from November and December 2023, while learning-based models are assessed from May-June 2020, the first two months following their training phase. Table~\ref{tab:ood_detection} confirms the trend we have seen in Table~\ref{tab:detection}: LAMD outperforms baselines and state-of-the-art methods even under this fair settings. We also observe notable tension between the two paradigms: the breakdown of F1 score reports a higher FPR and a lower FNR for LAMD compared to learning-based approaches. We hypothesize that zero-shot learners such as LAMD have better generalizability benefit from their large-scale pre-trained dataset that may generalize better against similar threats (resulting in lower FNR). Conversely, they suffer from making more mistakes due to the lack of a smaller, task-dependent dataset (lower FPR for learning-based approaches). If true, this opens interesting research opportunities to study tensions between the two paradigms.

\subsubsection{Effectiveness of explanations}
To assess analysis quality, we validate 100 correctly detected malware samples. 
Results show that 81 out of 100 samples are correctly classified into their respective categories. Table~\ref{tab:explanation} summarizes the sample distribution and classification accuracy across categories. Due to their less distinct malicious patterns, Adware and Riskware pose greater challenges to accurate analysis compared to other categories, contributing to their higher misclassification rates.


\begin{table}
\centering
\renewcommand{\arraystretch}{1.2} 
\caption{The quality evaluation of generated detection analysis. The ``Correct'' column indicates the number of categories correctly classified by the LLM.}
\label{tab:explanation}
\begin{tabular}{lccc}
\toprule
\textbf{Category} & \textbf{Family} & \textbf{Total} & \textbf{Correct} \\
\midrule
Adware    & gexin, ewind                     & 20 & 15 \\
Backdoor  & mobby, hiddad                    & 20 & 18 \\
PUA       & umpay, scamapp, apptrack         & 13 & 9  \\
Riskware  & jiagu, smspay, smsreg            & 25 & 18 \\
Scareware & fakeapp                          & 10 & 9  \\
Trojan    & hqwar, hypay                     & 12 & 12 \\
\midrule 
Overall   & /                                & 100 & 81 \\
\bottomrule
\end{tabular}
\end{table}

\subsubsection{Cost}
We estimate the overall cost of evaluating LAMD. Using GPT-4o-mini as our base LLM, which costs \$0.15 per 1M tokens for input and \$0.6 per 1M tokens for output~\cite{openai}, the total expense for our test set of 9,046 APKs amounts to approximately \$1800, averaging \$0.199 per apk. The cost for each apk varies depending on the number of sliced CFGs extracted and suspicious APIs the app invokes. While this demonstrates the feasibility of deploying LAMD in real-world scenarios, the operational expense remains non-trivial for large-scale applications. Future work could explore the effectiveness of applying LAMD with locally deployed LLMs like DeepSeek~\cite{deepseek} and Llama~\cite{llama} to avoid cloud-based API fees.

 \section{Case Study}
This section demonstrates how LAMD overcomes the limitations of current LLMs in Android malware detection. Since LLMs cannot process APKs directly, we use JADX\footnote{https://github.com/skylot/jadx} to decompile them, concatenating all pseudo source codes for input~\cite{llm4codeanalysis}. Besides GPT-4o-mini, we select Gemini 1.5 pro~\cite{gemini1.5} as another comparison, which claims their longest context windows and for its malware detection capabilities.


Figure~\ref{fig:motivation} highlights key failure cases. In the first example, we decompile a randomly selected sample\footnote{MD5: c37e223e3388b31b323ad39af45180fc}, which contains 1,547,806 lines of code, even after restricting the scope to the ``source/com'' folder containing critical files. Gemini 1.5 Pro fails to process it, exceeding its 20,971,520-byte context limit, while LAMD enables GPT-4o-mini (with 10× lower token capacity) to generate an accurate prediction, demonstrating its ability to efficiently analyze large-scale Android applications within constrained LLM contexts.


In the second case, we verify the detection ability of LAMD. Analyzing an \verb|SMSReg| malware sample\footnote{MD5: 2be97287c6af70f2074686b1a9021c06}, Gemini misclassifies it as ``BENIGN'', erroneously identifying it as part of the ``\textit{xUtils}” library due to excessive benign classes disguising malicious behavior. With LAMD, both GPT-4o-mini and Gemini 1.5 pro correctly classify it as malware and identify key behaviors. The \verb|SMSReg| malware family usually harvests device data, sends unauthorized SMS, and registers users for premium services to evading detection. Detection explanations enhanced by LAMD align with these behaviors, showing unauthorized access to \verb|getDeviceId()|, \verb|getSubscriberId()|, and \verb|sendTextMessage()|, reflection-based evasion, insecure SSL handling, and excessive location tracking—all indicative of SMS fraud and unauthorized transactions.


\begin{tcolorbox}[fontupper=\small, fontlower=\small, 
left=2pt, 
right=2pt, 
top=3pt, 
bottom=3pt, 
before upper=\raggedright, 
after upper={\vspace{-4pt}},
before lower=\raggedright]
\textbf{GPT-4o-mini Output with LAMD}:
\tcblower
\textbf{Final Prediction}: MALWARE \\
\textbf{Key Findings}: \\
* Sensitive API Misuse: Uses ``getDeviceId()'', ``getSubscriberId()'', and ``sendTextMessage()'' via reflection to bypass access controls for data harvesting.\\
* SSL Vulnerability: Unchecked ``SslErrorHandler.proceed()'' and ``onReceivedSslError'' expose users to MITM attacks.\\
* Location Tracking: Collects ``getLastKnownLocation'' and ``requestLocationUpdates'' without user consent.
\end{tcolorbox}

\begin{tcolorbox}[fontupper=\small, fontlower=\small, 
left=2pt, 
right=2pt, 
top=3pt, 
bottom=3pt, 
before upper=\raggedright, 
after upper={\vspace{-4pt}},
before lower=\raggedright]
\textbf{Gemini Output with LAMD}:
\tcblower
\textbf{Final Prediction}: MALWARE \\
\textbf{Key Findings}: \\
* Extensive use of sensitive APIs (e.g., ``getDeviceId()'', ``getSubscriberId()'', ``sendTextMessage()'') strongly suggesting data exfiltration and unauthorized actions. \\
* Unconditional ``SslErrorHandler.proceed()'' bypass risks data security and enables MITM attacks. \\
* Collects and transmits data to remote servers with obfuscation and poor error handling.\\
\end{tcolorbox}





\section{Discussion}
While LLMs improve Android malware detection, their general pre-training limits fine-grained behavior analysis. The phenomenon of over-squashing~\cite{barbero2024transformers} in LLMs still introduces challenges to successfully identify key aspects within complex programs. 
Future research can benefit from domain-specific fine-tuning or external knowledge integration~\cite{he2025benchmarking}. 
Meanwhile, learning-based methods remain essential for capturing overall insights in non-drift scenarios.
Our experiments find that while LLMs excel in adapting to unseen threats, they require significantly more computational resources and incur higher costs, highlighting effective strategies to combine these two paradigms and maximize their complementary strengths.
In future work, we plan to explore how LAMD and more generally context-aware LLMs perform against recent approaches aimed at learning temporally invariant features~\cite{zheng2025tif, he2024dream}.
These findings open interesting research opportunities to better understand the tension between learning-based models and pre-trained LLMs.
\section{Conclusion}
Large language models’ superior zero-shot inference offers a promising solution for Android malware detectors to handle distribution drift, dataset bias and explainability gaps in real-world scenarios but struggle with excessive support code and complex program structures. To address these challenges, we propose LAMD, the first practical framework enabling LLMs for explainable Android malware detection. Our evaluation in the real-world setting demonstrates that LAMD outperforms conventional detectors by effectively analyzing complex structures and semantics. LAMD unlocks LLMs' potential in Android security, paving the way for AI-driven malware analysis. 

\newpage
\bibliographystyle{IEEEtran}
\bibliography{reference}
\raggedbottom

\appendices
\section{Baseline Models}
\label{baseline}
Drebin~\cite{Arpdrebin} detects malware using binary feature vectors derived from nine data types (e.g., hardware components, API calls, permissions) and classifies samples via a linear classifier.

DeepDrebin~\cite{Grossedeepdrebin} extends Drebin by replacing the linear classifier with a three-layer deep neural network (DNN) while retaining the same feature space for feature extraction and classification.

Malscan~\cite{malscan} employs a graph-based approach, extracting sensitive API calls from APKs and computing four centrality measures (degree, Katz, proximity, and harmonic wave centralities) as features. We use the optimal feature with a Random Forest classifier, which leads to the lowest overhead.

All of them can be considered as the current state-of-the-art in Android malware detection.

\begin{table*}
\centering
\caption{Overview of the evaluation dataset, where M denotes malware and B denotes benign applications.}
\label{datasetdescription}
\setlength{\tabcolsep}{2.5pt}
\begin{tabular}{lcccccccc}
\toprule
\textbf{Test Set} & \textbf{Time Interval} & \textbf{Sample Size} & \textbf{Existing Families} & \textbf{New Families} & \textbf{Packed} & \textbf{Malicious (M)} & \textbf{Benign (B)} & \textbf{M/(M+B) \%} \\
\midrule
Test Set 1 & 2020.05 -- 2021.01 & 3015 & 21 & 24 & 18 & 284 & 2731 & 9.42 \\
Test Set 2 & 2021.01 -- 2021.12 & 3015 & 28 & 32 & 30 & 298 & 2717 & 9.88 \\
Test Set 3 & 2021.12 -- 2023.12 & 3016 & 34 & 36 & 40 & 302 & 2714 & 10.01 \\
\bottomrule
\end{tabular}
\end{table*}

\begin{table*}
\centering
\caption{Overview of the OOD evaluation dataset aligned with the GPT‑4o‑mini pre-training cutoff (October 2023), where M denotes malware and B denotes benign applications.}
\label{ood_datasetdescription}
\setlength{\tabcolsep}{1.7pt}
\begin{tabular}{lcccccccc}
\toprule
\textbf{Test Set} & \textbf{Time Interval} & \textbf{Sample Size} & \textbf{Existing Families} & \textbf{New Families} & \textbf{Packed} & \textbf{Malicious (M)} & \textbf{Benign (B)} & \textbf{M/(M+B) \%} \\
\midrule
Learning-based & 2020.05 -- 2020.06 & 884 & 15 & 7 & 2 & 62 & 822 & 7.01 \\
LAMD & 2023.11 -- 2023.12 & 60 & 0 & 1 & 0 & 7 & 53 & 11.67 \\
\bottomrule
\end{tabular}
\end{table*}


\begin{table*}
\centering
\caption{Performance of learning-based methods using Drebin features under different training/validation data volumes.}
\label{tab:dataselection}
\begin{tabular}{llcccccccccccc}
\toprule
\multirow{2}{*}{\textbf{Sample Size}} & \multirow{2}{*}{\textbf{Model}} 
& \multicolumn{3}{c}{\textbf{Validation}} 
& \multicolumn{3}{c}{\textbf{Test 1}} 
& \multicolumn{3}{c}{\textbf{Test 2}} 
& \multicolumn{3}{c}{\textbf{Test 3}} \\
\cmidrule(lr){3-5} \cmidrule(lr){6-8} \cmidrule(lr){9-11} \cmidrule(lr){12-14}
& & F1 & FPR & FNR & F1 & FPR & FNR & F1 & FPR & FNR & F1 & FPR & FNR \\
\midrule
\multirow{2}{*}{13k} 
& Drebin       & 94.02 & 0.57 & 6.58 & 81.33 & 0.40 & 24.21 & 73.60 & 0.99 & 31.14 & 61.59 & 4.36 & 37.00 \\
& DeepDrebin   & 95.72 & 0.24 & 6.06 & 71.92 & 0.62 & 34.12 & 69.22 & 0.85 & 36.13 & 66.11 & 0.95 & 38.58 \\
\midrule
\multirow{2}{*}{60k} 
& Drebin       & 95.83 & 0.34 & 5.28 & 68.76 & 0.68 & 36.67 & 63.41 & 1.91 & 40.12 & 62.83 & 2.33 & 40.28 \\
& DeepDrebin   & 96.68 & 0.28 & 4.15 & 68.82 & 0.64 & 36.65 & 63.48 & 1.72 & 40.17 & 58.99 & 4.31 & 41.19 \\
\bottomrule
\end{tabular}
\end{table*}

\section{Impact of Training Dataset Sizes}
\label{app:dataset}

In real-world scenarios, a large number of Android applications are packed and obfuscated to bypass malware detection systems and complicate manual analysis. A notable shift occurred post-2020, as more Androzoo samples adopted these techniques. Consequently, our test set, drawn from 2020, reflects a more realistic setting. 
Notably, the Jensen–Shannon~(JS) divergence~\cite{menendez1997jensen}~(measuring the similarity between two probability distributions) between the training and test sets is five times larger than that between the training and validation sets.

Intuitively, increased training samples should enhance detection performance in learning-based methods. While LAMD excels in zero-shot learning, one may argue that traditional methods are hindered by smaller training datasets compared to the vast pretraining data leveraged by LLMs. However, the impact of significant drift warrants reconsideration. Table~\ref{tab:dataselection} highlights the performance of learning-based detectors (e.g., Drebin feature space) across varying dataset sizes, across the training dataset timeframe.
The results indicate that additional training data does not improve detection of samples with significant drift, though it enhances performance on in-distribution samples. 

Expanding the training set introduces more information, helping detectors learn clearer decision boundaries, consistent with the principle of empirical risk minimization (ERM).
However, ERM identifies features that distinguish samples across the entire dataset, and when the training set is heavily skewed toward older data (e.g., 2014-2019/3), the model is more likely to learn features representative of early apps rather than more recent ones. Consequently, when tested on subsequent months (e.g., Test 1), the model exhibits worse performance, as it relies on features that are less effective for newer, drifted samples. This suggests that while increasing training data is generally beneficial, it may reinforce outdated patterns in the presence of significant drift, ultimately impairing generalizability.

\section{Factual Consistency Verification}
\label{datadependency}
\noindent To check factual consistency, we leverage data dependencies that capture relationships between variables and APIs. We focus on five dependencies (see Table~\ref{tab:data_dependency}) as they represent fundamental program relationships critical for malware reasoning and detection. Specifically:

\vspace{0.5em}
\noindent \hspace{0.5em} \textbf{(1) Variable-to-API Dependencies.} These dependencies determine whether the execution of an API is influenced by specific variables.


\begin{table*}[!t]
\centering
\caption{Data dependencies used for factual consistency verification.}
\label{tab:data_dependency}
\begin{tabular}{lll}
\toprule
\textbf{Category} & \textbf{Type} & \textbf{Description} \\
\midrule
\multirow{3}{*}{Variable-to-API} 
  & Direct      & A variable directly determines API execution. \\
  & Transitive  & A variable is propagated through a call chain to an API. \\
  & Conditional & A variable influences API execution via control flow (e.g., if/else). \\
\midrule
\multirow{2}{*}{Inter-variable} 
  & Parallel & Two variables jointly contribute to computing another variable. \\
  & Derived & A variable is derived from another through computation. \\
\bottomrule
\end{tabular}
\end{table*}

\begin{itemize}
    \item Direct dependencies: ensure that variables explicitly control API calls, providing strong evidence of intended execution.
    \item Transitive dependencies: track how variables propagate through function calls.
    \item Conditional dependencies: account for control-flow influences (e.g., if statements), identifying cases where malicious logic may be context-dependent.
\end{itemize}

\noindent \hspace{0.5em} \textbf{(2) Inter-variable Dependencies.} These capture relationships between variables that may affect security-sensitive operations.

\begin{itemize}
    \item Parallel dependencies: detect multiple variables jointly contributing to a computation, highlighting complex conditions leading to API execution.
    \item Derived dependencies: reveal computations that transform one variable into another, helping detect disguised or obfuscated malicious behaviors.
\end{itemize}



\noindent By incorporating these dependencies into factual consistency verification, LAMD ensures that summarized malicious behaviors are not based on hallucinatioms but are grounded in actual program logic, improving both precision and interpretability in Android malware detection.

\section{Backward Slicing Algorithm}
\label{slicing}

\noindent The proposed backward slicing algorithm is shown in Algorithm~\ref{alg1}. This backward slicing procedure consists of two phases: variable retrieval and slice extraction. In Stage 1, the algorithm tracks variables that influence the suspicious API invocation $a_i$. Specifically, in Line 2, the algorithm initializes the $worklist$ with $unit$ containing $a_i$, and in Line 4, it begins iterating through the control flow graph (CFG) in a backward manner starting from $unit$, extracting relevant variables from each unit. The algorithm updates the locals list by removing defined variables and adding those that affect $a_i$ in Line 13, and stores the updated variables in $varMap$ in Line 15. Stage 2 focuses on extracting slices based on $varMap$. In Line 23, the algorithm resets the $worklist$ and $visited$ sets, then starts from the predecessors of the original unit that invoked $a_i$. Line 28 checks the successors of each unit and identifies those relevant to the API invocation. The combined slices are formed in Line 29, integrating all the relevant variables and units, and the algorithm returns the slices in Line 38. This two-phase process enables the algorithm to isolate the parts of the function that influence the suspicious API invocation, facilitating further analysis.

\begin{algorithm}
\caption{Algorithm of Backward Slicing}
\label{alg1}
\begin{algorithmic}[1]

\REQUIRE ~Control Flow Graph $CFG$ of the function that invokes the suspicious API $a_i$ at unit block $unit$
\STATE \textbf{Stage 1: Variable Retrieval}
\STATE Initialize $worklist$ with $unit$ containing $a_i$
\STATE Initialize $visited \gets \{\}$, $varMap \gets \emptyset$

\WHILE {$worklist \neq \emptyset$}
    \STATE Pop $currUnit$ from $worklist$
    \STATE Initialize $locals \gets \{\}$

    \IF {$currUnit$ contains $a_i$}
        \STATE Extract variables used by $a_i$ and add to $locals$
    \ELSE
        \STATE Load variables from successors of $currUnit$ using $varMap$ and add to $locals$
    \ENDIF

    \IF {$currUnit$ is relevant to the $a_i$ invocation based on $locals$}
        \STATE Update $locals$: remove defined variables, add used variables that affect the $a_i$ invocation
    \ENDIF

    \STATE Store $locals$ in $varMap[currUnit]$

    \FOR {$pred \in$ predecessors of $currUnit$}
        \IF {$pred \notin visited$}
            \STATE Add predecessor $pred$ to $visited$ and $worklist$
        \ENDIF
    \ENDFOR
\ENDWHILE

\STATE \textbf{Stage 2: Slices Extraction}
\STATE Initialize $slices \gets \{unit\}$, reset $worklist$ and $visited$
\STATE Add predecessors of $unit$ to $worklist$

\WHILE {$worklist \neq \emptyset$}
    \STATE Pop $currUnit$ from $worklist$
    \STATE Load variables from the successors of $currUnit$ using $varMap$

    \IF {$currUnit$ is relevant to the $a_i$ invocation based on the successor variable}
        \STATE Add $currUnit$ to $slices$
    \ENDIF

     \FOR {$pred \in$ predecessors of $currUnit$}
        \IF {$pred \notin visited$}
            \STATE Add predecessor $pred$ to $visited$ and $worklist$
        \ENDIF
    \ENDFOR
\ENDWHILE

\STATE \RETURN $slices$

\end{algorithmic}
\end{algorithm}

%


\end{document}